\documentclass[a4paper,12pt]{article}

\usepackage{graphicx}
\usepackage{amsmath}
\usepackage{amsfonts}
\usepackage{amssymb}
\usepackage{units}
\usepackage{epsf}
\usepackage{bbm}   
\newcommand{\be}{\begin{equation}}
\newcommand{\ee}{\end{equation}}
\newcommand{\ba}{\begin{align}}
\newcommand{\ea}{\end{align}}

\newcommand{\ket}[1]{\ensuremath{| \, #1 \, \rangle }}

\newcommand{\1}{\ensuremath{\mathbbm{1}}}

\begin{document}
\title{Flow of entropy in the evolution of the $B^0-\bar{B}^0$ system:
Upper bound on $CP$ violation from unidirectionality
\author{Ch. Berger \thanks{e-mail: berger@rwth-aachen.de}
\\{\small I. Physikalisches Institut, RWTH Aachen University, Germany} \and
L. M. Sehgal\thanks{e-mail: sehgal@physik.rwth-aachen.de}
\\{\small Institute for Theoretical Particle Physics and Cosmology,}
\\{\small RWTH Aachen University, Germany}}
}

\date{}

\maketitle

\begin{abstract}
We have previously studied~\cite{CPVpaper,BS2} the time-dependence of a $B^0-\bar{B}^0$
mixture in terms of its density matrix 
$\rho(t) = N(t) (\1 + \vec{\zeta}(t) \cdot \vec{\sigma})/2$.
The requirement that $\zeta(t)$, the absolute value of the Stokes vector $\vec\zeta(t)$,
 should evolve monotonically from its
initial value $\zeta(0)=0$ to its final
value $\zeta(\infty)=1$ was shown to lead to an upper bound on the $CP$ violating overlap
$\delta=\langle B_L|B_S\rangle$.
In the present note, we consider the entropy variable 
$S=-{\rm tr}(\tilde\varrho\log_2(\tilde\varrho))$, where $\tilde\varrho=\varrho/N$, as an alternative
measure of mixing.
We show that exactly the same
upper bound emerges from the requirement that the flow of entropy is unidirectional ($dS/dt<0$).
We compare the entropic current $dS/dt$ with and without
$CP$ violation and identify certain physical features that appear when the bound on $\delta$ is violated.

\end{abstract}

\section{Evolution of $B^0-\bar{B}^0$ in terms of Stokes vector}
In a previous paper~\cite{CPVpaper} we studied the manner in which a $B^0-\bar{B}^0$
state, prepared as an equal incoherent mixture, evolves into a final 
coherent (pure) state representing the long-lived neutral
meson $B_L$. The analysis was done in terms of the 2 x 2 density matrix
\be
 \rho(t) = \frac{1}{2} N(t) \left[\1 + \vec{\zeta}(t) \cdot \vec{\sigma} \right]
  \label{CPV3}
 \ee
with initial values $N(0)=1$, $\zeta(0)=|\vec\zeta(0)|=0.$ Explicit solution of
the Schr{\"o}dinger equation for $\ket{B^0}$ and $\ket{\bar B^0}$ yields 
the result
\be N(t) = \frac{1}{2(1-\delta^2)} \left[ e^{- t} + e^{-
r t} - 2 \delta^2 e^{ - \frac{1}{2} ( 1+r)t} \cos \mu t \right]\label{CPV13a}
\ee
and
\be
\zeta(t)= \left[1- \frac{1}{N(t)^2} e^{- ( 1+r)t}\right]^{\frac{1}{2}}\enspace .
\ee
Here $t$ is the proper time measured in units of $\tau_S$, the lifetime of the short-lived
eigenstate. In addition $r=\gamma_L/\gamma_S$, $\mu=\Delta m/\gamma_S$ with $\Delta m=
m_L-m_S$. The parameter $\delta$ is the $CP$ violating overlap of the two eigenstates
\be
\delta=\langle B_L|B_S\rangle\enspace .
\ee
It was shown that $\zeta(t)$, the magnitude of the Stokes vector, undergoes
a transition from monotonic to nonmonotonic
behaviour at a critical value
\be
\delta_{\rm crit} = \sqrt{\frac{1}{2} \left( \frac{1- r}{\mu}\right)
 \sinh \left( \frac{3\pi}{4} \; \frac{\left( 1- r\right) }{\mu}\right)}
\approx\sqrt{\frac{3\pi}{8}}\frac{1-r}{\mu}\enspace .
\label{CPV24}
\ee
For the $B_s^0-\bar{B}_s^0$ system, the theoretically expected value of $\Delta\gamma=\gamma_S-\gamma_L$
is
$0.087\pm 0.021\,\, {\rm ps}^{-1}$~\cite{Lenz2}, compatible with the experimental value 
$0.100\pm 0.013\,\, {\rm ps}^{-1}$ as evaluated by the HFAG group~\cite{HFAG}.
From the same source we  get $\Delta m=17.69\pm 0.08\,\, {\rm ps}^{-1}$.
With $r=0.85\pm 0.03$ and $\mu=26.4\pm 0.18$  the critical value of the $CP$ violating
parameter in (\ref{CPV24}) is 
$\delta_{\rm crit}=(0.62\pm 0.02)\%$. This is nearly three orders of magnitude higher
than the expected value $\delta\approx 1.0\times 10^{-5}$ in 
the standard model~\cite{Lenz}. Surprisingly, an empirical determination of
$\delta$ for the  $B_s^0-\bar{B}_s^0$ system, based on an asymmetry between $\mu^+\mu^+$ and
$\mu^-\mu^-$ events in the D0 experiment~\cite{D0} had given a value incompatible with the standard model
at the $2\sigma$ level. The recent evaluation of all experimental results
by the HFAG group~\cite{HFAG} yields an average
$\delta_{\rm exp}=(0.52\pm0.32)\%$ now compatible
with the standard model prediction and the limit obtained from (\ref{CPV24}).
The
unexpected D0 result and the intensive discussion in the
literature about its consequences, see e.g.~\cite{BS2,Review}, induced us to examine further
the implication of
the phase transition pointed out in~\cite{CPVpaper}.

\section{Evolution of $B^0-\bar{B}^0$ in terms of entropy}
To obtain further insight into the nature of the phase transition in $\zeta(t)$, we introduce 
a new variable $S(t)$ which is connected to the Stokes vector via
\be
S(t)=-\frac{1+\zeta(t)}{2}\log_2\left(\frac{1+\zeta(t)}{2}\right)-
\frac{1-\zeta(t)}{2}\log_2\left(\frac{1-\zeta(t)}{2}\right)
\enspace .
\label{entropdef0}\ee
This may be written formally as 
\be 
S=-{\rm tr}(\tilde\varrho\log_2(\tilde\varrho))=
-\tilde\lambda_1\log_2(\tilde\lambda_1) -\tilde\lambda_2\log_2(\tilde\lambda_2)
\label{entropdef} \ee
where $\tilde\varrho=\varrho/N$ and $\tilde\lambda_1,\tilde\lambda_2$ are the eigenvalues of
$\tilde\varrho$, given explicitly by
\be
\tilde\lambda_1=\frac{1}{2}(1+\zeta),\enspace \enspace \tilde\lambda_2=\frac{1}{2}(1-\zeta)\enspace .
\ee
The function  $S(t)$ is closely related to the  traditional von Neumann entropy 
$S_{\rm vN}=-{\rm tr}(\varrho\log_2(\varrho))$ defined for stable systems with $N(t)=tr(\varrho)=1$.
The definition (\ref{entropdef}) takes into account that for a decaying  $B^0-\bar{B}^0$ system
the norm $N(t)$ is different from unity.

The entropy as defined in (\ref{entropdef}) has the following desirable properties
\begin{enumerate}
 \item[(i)] $S(t)$ lies between 0 and 1 for all $\delta,r,\mu,t$.
\item[(ii)] For $\zeta=0$ (incoherent limit) , $S=1$ (maximum disorder). For $\zeta=1$ (coherent limit),
$S=0$ (minimum disorder).
\item[(iii)] $S(t)$ has the important feature that the derivative is related to the derivative
of $\zeta(t)$ by
\be 
\frac{dS}{dt}=-\frac{1}{2}\log_2\left(\frac{1+\zeta}{1-\zeta×}\right)\frac{d\zeta}{dt}\enspace .
\ee
\end{enumerate}
This shows that the zeros of $dS/dt$ are the same as the zeros of $d\zeta/dt$, implying
that the critical value of the $CP$ violating parameter at which  $dS/dt$ and $d\zeta/dt$
change sign is the same.
\begin{figure}[h]
\begin{center}
\includegraphics[width=0.74\textwidth]{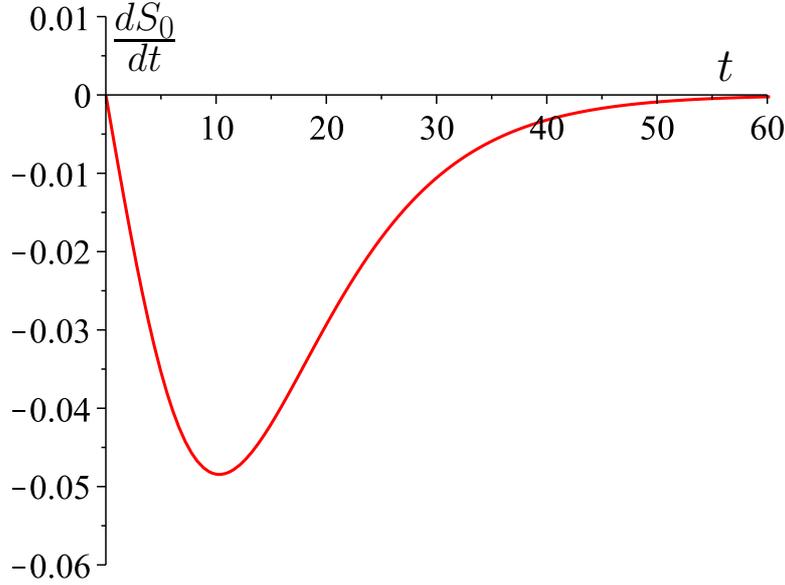}
\end{center}

\caption{{\small $dS_0/dt$, the entropic current for a  $B_s^0-\bar{B}_s^0$ like system 
in the tranquil state ($\delta=0$) versus
$t$ in units of $\tau_S$, the lifetime of $B^0_S$. The area under the curve is $-1$.
}
}
\end{figure}

\section{Entropic current in the absence of $CP$ violation}
We first consider the entropy function $S(t)=S_0(t)$ when there is no $CP$ violation ($\delta=0)$.
In this limit the modulus of the Stokes vector to be inserted in  (\ref{entropdef0}) is given by
\be
\zeta_0(t)=\frac{e^{-t}-e^{-rt}}{e^{-t}+e^{-rt}}\enspace .
\ee

An important characteristic is the derivative $dS_0/dt$ which is
given by
\be
\frac{dS_0}{dt}=\frac{-b^2te^{-bt}}{\ln{(2)}(1+e^{-bt})^2}\enspace .
\ee
This function is entirely determined by the parameter $b=1-r$, with no dependence on $\mu$. It
is plotted in fig.1 behaving as $-t\exp(-bt)$ at large $t$ and $-b^2t/(4\ln{2})$ at small $t$.
The negative sign reflects the fact that $S_0$ is a decreasing function of time, for all $t$. 
The decrease of entropy in the 
present situation contrasts with the increase of entropy familiar from stable thermodynamical systems.
It is a consequence of the fact that a mixed  $B^0-\bar{B}^0$ state, with arbitrary $S_0(0)>0$
 must ultimately reduce to the pure
long-lived state $B_L$ when the short-lived component has died out, implying that $S_0(\infty) =0$.
An interesting
feature is the location of the minimum which is calculated from the  the root of
the transcendental equation
\be
e^{-bt}=\frac{bt-1}{bt+1}
\ee
The numerical result is 
\be
t=\frac{1.543405}{1-r}\enspace .
\ee
We will refer to the curve $dS_0/dt$ as the ``entropic current in the tranquil state'' ($\delta=0$). 
 
\begin{figure}[h]
\begin{center}
\includegraphics[width=0.70\textwidth]{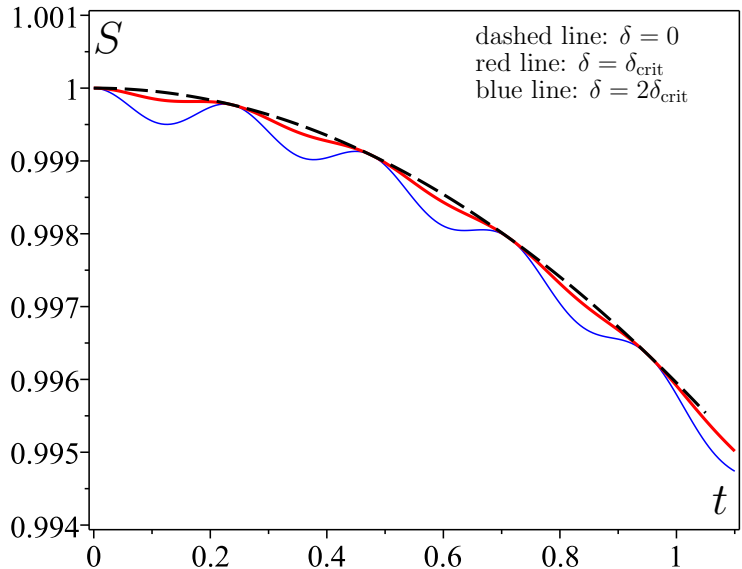}
\end{center}

\caption{{\small Entropy function $S(t)$ versus $t\leq 1$ for a  $B_s^0-\bar{B}_s^0$ like system
using $\delta=0,\delta_{\rm crit},2 \delta_{\rm crit}$, For large values of $t$ the oscillations
are strongly damped. The curve for $\delta=\delta_{\rm crit}$ defines the boundary between monotonic 
and nonmonotonic behaviour of $\zeta(t)$.
}
}
\end{figure}

\section{Impact of $CP$ violation on entropic current}
When $CP$ violation is taken into account ($\delta\neq 0$) the modulus of the Stokes vector  changes to 
\be
\zeta(t)= \sqrt{1-\frac{e^{-(1+r)t}}{N(t)^2}}
\ee
where $N(t)$ is given in eq.(\ref{CPV13a}). Inserting this $\zeta(t)$ in (\ref{entropdef0}) 
yields $S(t)$ as shown in fig.2 for $\delta=\delta_{\rm crit}$ and  $\delta=2\delta_{\rm crit}$.
For comparison $S_0(t)$ is also included. 

The impact of $CP$ violation is more clearly seen in the entropic current $dS/dt$ and its comparison 
with $dS_0/dt$. This comparison is shown in fig.3. The effect of $CP$ violation is essentially
a modulation of the curve $dS_0/dt$ by oscillations.
The curve $dS_0/dt$ is, in the interval $0<t<2$, practically a straight line. 
The modulation consists of an oscillating function, with frequency $\Delta m$ and amplitude
proportional to $\delta^2$. For $\delta<\delta_{\rm crit}$ the ampltude of the oscillations
(``ripples'') is small enough that the entropic current remains negative in sign ($dS/dt<0$). 
For  $\delta >\delta_{\rm crit}$ the oscillations grow in amplitude to the extent
that some of the ripples become ``eddies'': these are regions in which the entropic 
current reverses its sign. This fact is illustrated in fig.3 with the help of two curves 
showing the entropic current for a value $\delta=0.0055$ slightly below $\delta_{\rm crit}$ and
$\delta=2 \delta_{\rm crit}$ respectively. The transition in the sign of $dS/dt$ as
$\delta$ crosses the critical value $\delta_{\rm crit}$  is equivalent to the transition
from monotonic to nonmonotonic behaviour of  the Stokes vector $\zeta(t)$
noted  in~\cite{CPVpaper}.
 
For completeness the effect of $CP$ violation at larger values of $t$ especially in 
the neighborhood of the minimum of fig.1 is shown in fig.4.

\begin{figure}[h]
\begin{center}
\includegraphics[width=0.74\textwidth]{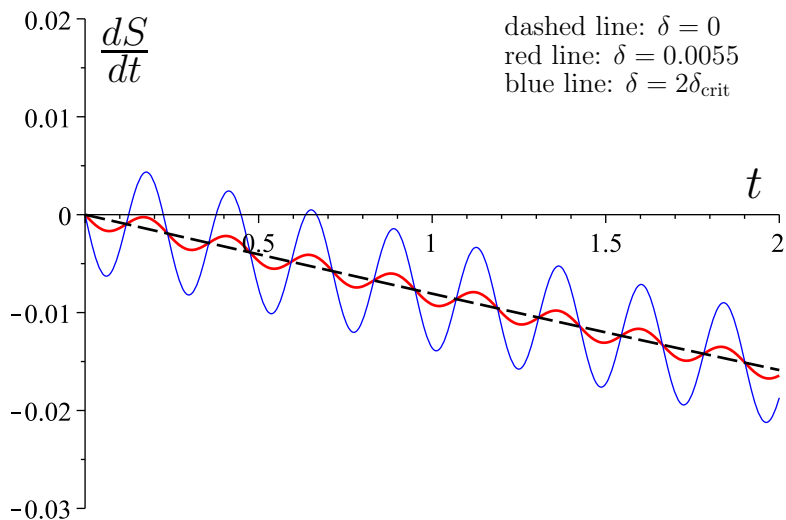}
\end{center}

\caption{{\small Entropic current $dS(t)/dt$ for a  $B_s^0-\bar{B}_s^0$ 
like system versus $t\leq 2$ 
for $\delta=0$ (straight line), $\delta <\delta_{\rm crit}$ (ripples) and $\delta =2\delta_{\rm crit}$
(eddies).
}
}
\end{figure}
\begin{figure}[h]
\begin{center}
\includegraphics[width=0.80\textwidth]{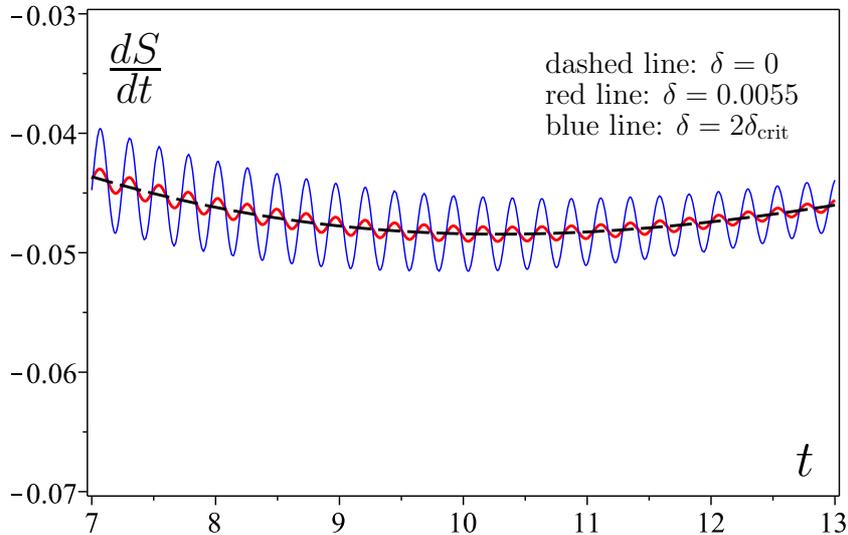}
\end{center}

\caption{{\small Entropic current $dS(t)/dt$ for a  $B_s^0-\bar{B}_s^0$ like system
versus $t$ in the neighborhood 
of the minimum of fig.1
for $\delta=0$ (dashed curve), showing small ripples for $\delta <\delta_{\rm crit}$ and 
larger ripples for $\delta =2\delta_{\rm crit}$.
}
}
\end{figure}

\section{Remarks}
\begin{enumerate}

\item[(i)] Our purpose in this note has been to elucidate the phase transition in the evolution of
a $B_0-\bar B_0$ mixture, pointed out in Ref.~\cite{CPVpaper}, by going from a Stokes vector description 
to one 
involving entropy. There is a change in sign in the entropic current $dS/dt$
at exactly the same 
critical value of the $CP$ violating parameter $\delta$, written explicitly in eq.~(\ref{CPV24}).
Examination of the entropy flow below and above the
critical value (fig.3) shows that the loss of unidirectionality of the entropic flow is 
accompanied by the appearance of one or more eddy-like
transients that cause a reversal in the sign of $dS/dt$ in limited intervals of time.

\item[(ii)] The onset of $CP$ violation causes,in the subcritical domain,
a ripple-like perturbation in $dS/dt$, with an amplitude proportional to
$\delta^2$, which is not large enough to change the negative sign of the entropic current.
For  $\delta>\delta_{\rm crit}$ these ripples grow in size to the 
extent that the entropic current penetrates into the region $dS/dt>0$. 
When that happens,the arrow of time reverses direction in certain
intervals of time, compared to its direction in the tranquil state.

\item[(iii)] Our study has focussed on a simple two-level system whose eigenstates have different masses 
and lifetimes. In the absence of $CP$ violation, the quantum mechanical entropy of such an unstable (open) 
system decreases with time. We have found that this behaviour is affected by $CP$ violation.
In a recent paper~\cite{Goldman} the question has been raised whether the monotonic
 increase of entropy in the macroscopic 
universe (the progression from order to disorder) can also be influenced by $CP$ violation.
Examples are given where 
$CP$ or $T$ violation can reverse this evolution. In a different context~\cite{Vaccaro}
 it has been argued that the 
unidirectionality of time could itself be a consequence of $CP$ violation. Our analysis of
the $B_0-\bar B_0$ system shows that 
questions concerning entropy and the impact of $T$-violation on the arrow of time
can be discussed in a meaningful way also in the context of pristine two-level systems such as
$K_0-\bar K_0$ or $B_0-\bar B_0$,
which have given us profound insights into the nature of the world under $C, P$, and $T$ transformations.

\end{enumerate}


\begin{thebibliography}{xx}
\bibitem{CPVpaper} Ch. Berger and L.M. Sehgal, Phys. Rev. D {\bf 76}, 036003 (2007);
\\ arXiv:0704.1232v2 [hep-ph]
\bibitem{Lenz2} A.Lenz and U. Nierste, arXiv: 1102.4274v1 [hep-ph]
\bibitem{Lenz} A.Lenz, arXiv: 1205.1444v1 [hep-ph]
%
%
\bibitem{HFAG} HFAG Group, Results for the PDG 2012 review, to be found online under
http://www.slac.stanford.edu/xorg/hfag/osc/PDG\_2012/\#CPV
\bibitem{D0} V.M. Abazov et al. (D0 Collaboration),
Phys. Rev. D {\bf 82}, 03201 (2010), arXiv:1005.2757v1 [hep-ex];
V.M. Abazov et al. (D0 Collaboration), Phys. Rev. Lett. {\bf 105}, 081801 (2010), arXiv:1007.0395v1 [hep-ex]
\bibitem{BS2} Ch.Berger and L.M.Sehgal, Phys.Rev. D {\bf 83}, 037901 (2011); arXiv: 1007.2996v3 [hep-ph]
\bibitem{Review} See, for example, M. Freytsis, Z. Ligeti and S. Turczyk, 
arXiv: 1203.3545v2 [hep-ph] and references therein
\bibitem{Goldman} T.Goldman and D.H.Sharp, Europhys. Lett. {\bf 97} (2012) 61003; arXiv:1203.6092v1 [hep-ph]
\bibitem{Vaccaro}  J.Vaccaro, Found.Phys. {\bf 41} (2011) 1569; arXiv:0911.4528v3 [quant-ph]



\end{thebibliography}
\end{document}